\documentclass{PoS}

\usepackage{url}
\usepackage{bm}
\usepackage{graphicx}
\usepackage{epsfig,times,color,cite,adjustbox,caption,wrapfig,lipsum,booktabs,comment}

\title{MAGIC sensitivity to millisecond-duration optical pulses}
\ShortTitle{MAGIC sensitivity to short optical pulses}

\author{\speaker{T. Hassan}$^{1}$, J. Hoang$^{2}$, M. L\'{o}pez$^{2}$, J. A. Barrio$^{2}$, J. Cortina$^{1}$, D. Fidalgo$^{2}$, D. Fink$^{3}$, L. A. Tejedor$^{2}$, M. Will$^{3}$\\
       $^{1}$Institut de Fisica d'Altes Energies (IFAE), The Barcelona Institute of Science and Technology\\
       $^{2}$Universidad Complutense de Madrid, Grupo de Altas Energ\'{i}as (GAE)\\
       $^{3}$Max Planck	Institute for Physics\\
       E-mail: \email{thassan@ifae.es}}

\abstract{The MAGIC telescopes are a system of two Imaging Atmospheric Cherenkov Telescopes (IACTs) designed to observe very high energy (VHE) gamma rays above ~50 GeV. However, as IACTs are sensitive to Cherenkov light in the UV/blue and use photo-detectors with a time response well below the ms scale, MAGIC is also able to perform simultaneous optical observations. Through an alternative system installed in the central PMT of MAGIC II camera, the so-called central pixel, MAGIC is sensitive to short (1ms - 1s) optical pulses. Periodic signals from the Crab pulsar are regularly monitored. Here we report for the first time the experimental determination of the sensitivity of the central pixel to isolated 1-10 ms long optical pulses. The result of this study is relevant for searches of fast transients such as Fast Radio Bursts (FRBs).}

\FullConference{35th International Cosmic Ray Conference $–$ ICRC2017$-$\\
		10$-$20 July, 2017\\
		Bexco, Busan, Korea}

\begin{document}

\section{Introduction}

The discovery of new very fast transient phenomena such as Fast Radio Bursts (FRBs) \cite{FRB_discov, FRB_repeater} motivates the development of new detection techniques to search for brief electromagnetic bursts all along the multi-wavelength spectrum \cite{aqueye, aqueye+, ultracam, quanteye}. Imaging Atmospheric Cherenkov Telescopes (IACTs) provide excellent sensitivity for transient phenomena in the Very High Energy (VHE) gamma-ray regime (E > 50 GeV) mainly due to their large collection area. These telescopes are sensitive to the optical Cherenkov flashes produced by the effect of charged secondary particles generated within extensive air showers. In addition, their large reflecting surface and isochronicity makes them suitable detectors to detect faint short (1 ms -- 1 s) optical pulses \cite{MAGIC_cpix, veritas_opt, hess}.

The central pixel of MAGIC-II telescope allows simultaneous observations within both VHE and optical energy ranges. This system will be particularly beneficial for searching for fast optical variable sources as well as other transient phenomena such as FRBs. This device has been classically used to measure the Crab Pulsar light curve \cite{MAGIC_cpix} but here we also explore its sensitivity for orphan optical flashes.




\section{The upgraded MAGIC Central Pixel System}
\label{sec:cpix}


The MAGIC central pixel system consists of a fully modified photosensor-to-readout chain at the center of the MAGIC-II telescope camera~\cite{MAGIC_II}. A standard MAGIC pixel comprises of a PMT followed by a preamplifier step that splits the signal in two branches: a so-called \emph{AC branch}, which processes Cherenkov light pulses with a high bandwidth, and a \emph{DC branch}, monitoring the PMT anode current. The central pixel system involves modifying such DC branch in order to increase its  bandwidth from the 8 Hz of a normal DC branch to over 3 kHz, which dominates the bandwidth of the whole system. That new DC branch is fed to both the standard DC monitoring system and to an additional optical transmitter that sends the signal down to the Control house. Once there, it is converted and adapted to be read-out both by standard MAGIC DAQ system (dubbed hereon \emph{Domino readout}), and also delivered to the  central pixel PC for a dedicated readout. In what follows, a brief summary of the MAGIC central pixel system is presented.

\begin{figure}[ht]
\centering
\includegraphics[width=0.6\textwidth]{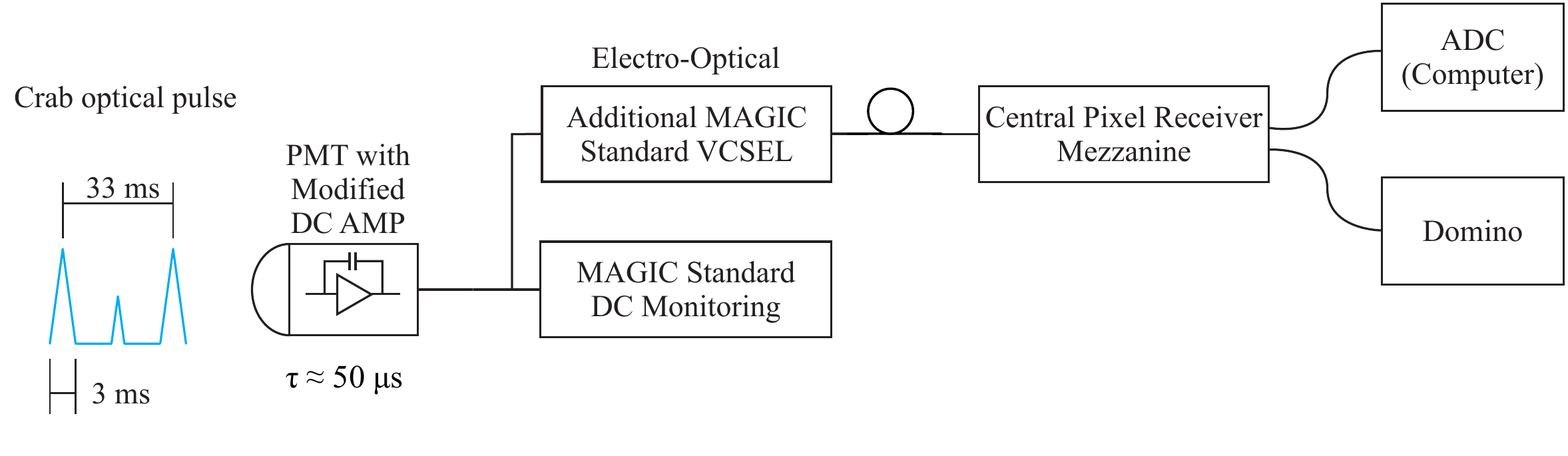}
\caption{\it  central pixel system block diagram, showing the upgraded Central Pixel DC branch. The central pixel AC branch, used to detect pulses of Cherenkov photons associated to $\gamma$-ray-induced showers, is not drawn in the diagram.}
\label{fig:CentralPixelBlockDiagram}
\end{figure}

The old camera of MAGIC-I telescope, prior to the MAGIC upgrade~\cite{MII_camera}, was also incorporating a central pixel system~\cite{MAGIC_cpix} that was instrumental in the first detection of the Crab pulsar in the VHE regime by the MAGIC telescopes~\cite{MAGIC_CP}. After the above-mentioned upgrade, the central pixel system was also fully refurbished and installed in MAGIC-II telescope camera.

\subsection{The Central Pixel Photomultiplier and preamplifier}

The PMTs installed in MAGIC have attached to them several electronic boards, one of them implementing the preamplifier system, which include both the AC and DC branches. In order to increase the bandwidth of the DC branch up to few kHz, a few components of the preamplifier had to be exchanged. Care was taken so that the AC branch high-bandwidth was preserved, so that the central pixel can also operate as a standard MAGIC pixel. 


\subsection{The Central Pixel signal transmission system}

The standard MAGIC DC-branch monitoring system involves low-pass filter of a few Hz. Therefore, in order to preserve the required few kHz bandwidth, a dedicated signal transmission system has been implemented, which includes an optical transmitter at the MAGIC camera, the optical fiber transporting the signal down to the MAGIC counting house and an optical receiver at the counting house itself.   

\subsection*{The Optical Transmitter}

\begin{figure}[ht]
\centering
\includegraphics[width=0.8\textwidth]{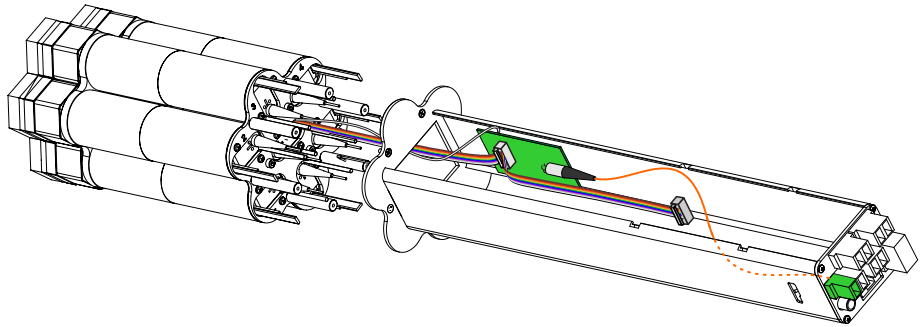}
\caption{\it MAGIC cluster hosting the central pixel, including the location of the  optical transmitter board, in green. The VCSEL laser for the central pixel, and its corresponding optical fiber (in orange) are also highlighted.}
\label{fig:opticalTransmitterLocation}
\end{figure}

The central pixel signal uses one MAGIC optical fibers to deliver its slow (few kHz) signal to the  counting house. In order to achieve such optical transmission, a dedicated electrical to optical converter has been developed, adapted to the needs of the special central pixel optical receiver. Basically, it contains a single stage consisting of an operational amplifier in a non-inverting configuration driving a VCSEL laser. 
In Figure~\ref{fig:opticalTransmitterLocation}, the board implementing the optical transmitter is shown inside the MAGIC cluster hosting the central pixel. Also in green, it is shown the special socket to connect the optical fiber that transports the central pixel optical signal.   

\subsection*{The Optical Receiver mezzanine}
\label{sec:mezz}
In order to process the optical signal that comes from the camera to the counting house for its digitalization, a dedicated receiver circuit is needed. The low frequency signal characteristics of the central pixel output (few kHz) makes it impossible to use the standard MAGIC receiver analog channels, involving typical high-pass filters of several MHz. Thus, one channel in a MAGIC receiver board is \emph{replaced} by the central pixel Optical receiver, implemented in a mezzanine board, which is connected to that MAGIC Receiver board. 
In this mezzanine, the optical signal is converted back to electrical, conditioned, split so that it is sent both (Fig. \ref{fig:CentralPixelBlockDiagram}) to a Domino readout channel  and to the  central pixel PC via a LEMO bipolar connector (described in detail in~\ref{sec:pc-daq}). 



\subsection{The Central Pixel digitizing system}
\label{sec:pc-daq}
In order to achieve the maximum sensitivity for the central pixel system, its signals are digitized at the MAGIC counting house in a dedicated way, independent of the MAGIC standard readout system. Therefore, the differential signal produced at the Optical receiver mezzanine is  digitized  by a National Instrument PCIe 6251 M Series ADC card, with 16-bit resolution. The ADC card is hosted in a dedicated computer, the so-called central pixel PC, that will also store the central pixel digitized signals produced by the ADC. The ADC samples the central pixel signal at a 10 kHz rate, using an external signal provided by th e Rubidium clock Oscillator of the MAGIC Timing system.


It is worth mentioning, as it has been described in~\ref{sec:mezz}, that the central pixel signal is also digitized in a standard MAGIC Domino channel. In this way, the Optical Crab pulsation can also be detected by the standard MAGIC DAQ, although with lower sensitivity than with the dedicated ADC channel. However, this detection allows to verify the MAGIC time-stamping system against the most precise clock the nature can provide.


\section{Central pixel sensitivity from periodic optical pulses: Crab Pulsar}
\label{sec:crab}

\begin{figure}[ht]
\centering
\includegraphics[width=0.7\textwidth]{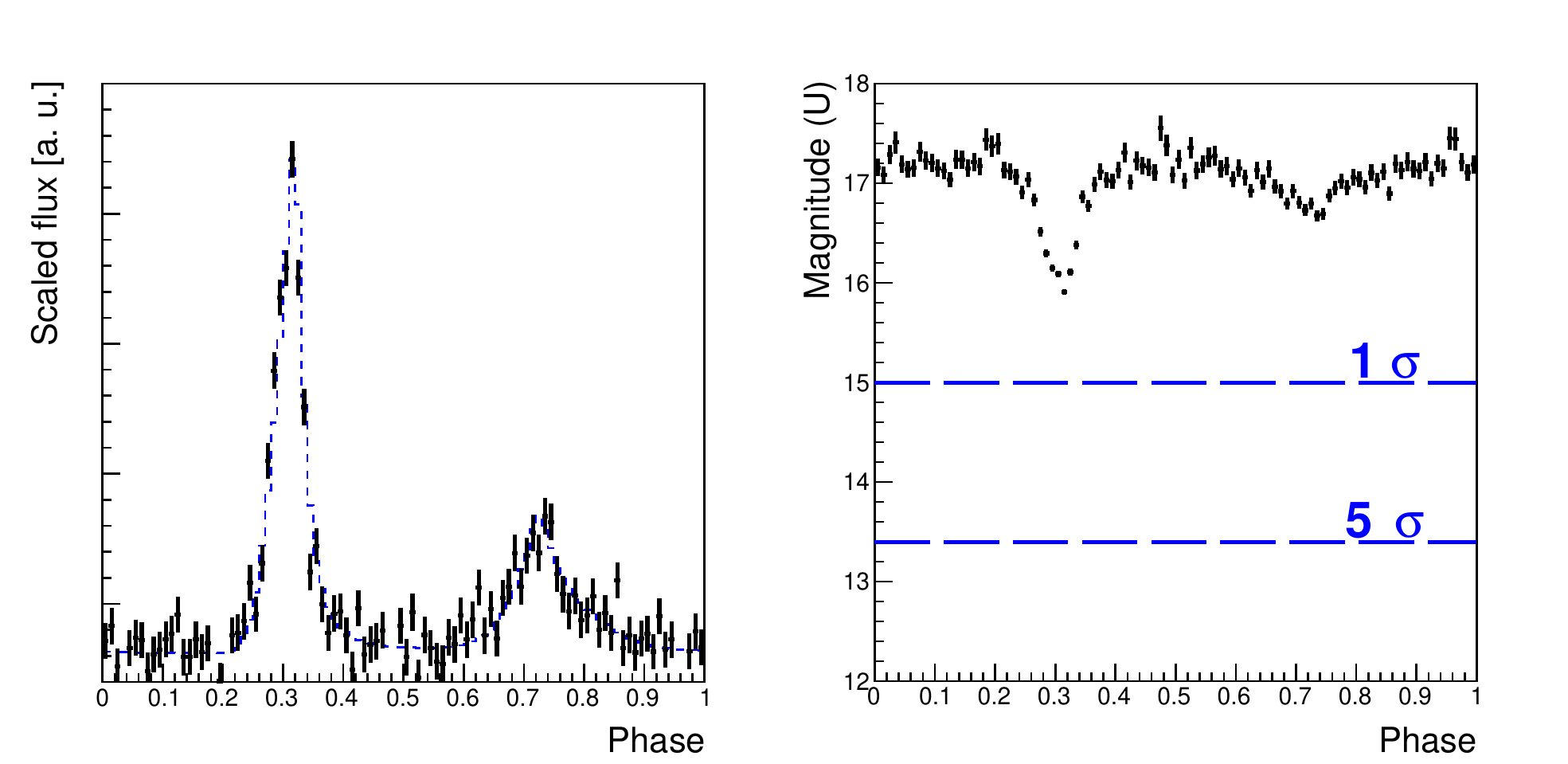}
\caption{\textit{Left}: Crab light-curve after 5 minutes of observation with the MAGIC central pixel, compared with an equally binned light-curve taken by Aqueye \cite{aqueye}. \textit{Right}: U band magnitude of the same Crab pulsar light-curve, as compared with the equivalent flux levels of 1 and 5 $\sigma$ deviations from the background noise of the central pixel. To reduce the high frequency noise, a 1 ms averaging filter is applied to the background.}
\label{fig:Folded Crab}
\end{figure}

As previously demonstrated, MAGIC is capable of detecting the Crab pulsation in very short observation times \cite{MAGIC_CP, MAGIC_cpix}. Left panel of Fig. \ref{fig:Folded Crab} shows the folded light-curve of the Crab Pulsar resulting from 5 minutes of central pixel data, compared with the one obtained recently by Aqueye \cite{aqueye}. This light-curve was obtained using an equivalent analysis as in \cite{MAGIC_cpix}, folding the absolute times of each sample with the expected Crab period, in this case using Jodrell Bank observatory\footnote{http://www.jb.man.ac.uk/pulsar/crab.html} radio ephemeris. After the upgrade described in section \ref{sec:cpix}, the required time for detecting the Crab pulsation has been reduced down to less than 10 s.

As a first attempt to estimate the sensitivity of the MAGIC central pixel to orphan ms pulses, we compare the RMS noise with the well understood Crab Pulsar light-curve. As the central pixel data are affected by higher frequency noise (mainly caused by the frequency of the power supply) a 1 ms averaging filter is applied to estimate the RMS of the background by using off-source data. By using the average flux of the Crab Pulsar within the U band \cite{Bessel,Sollerman}, and assuming a pulse shape equivalent to the (finely binned) Crab light-curve taken by Aqueye \cite{aqueye}, we convert the voltage of the measured light-curve to magnitudes in the U band. As shown in the right panel of Fig. \ref{fig:Folded Crab}, Crab pulses are well below the 1$\sigma$ level of our background noise, showing the large sensitivity improvement produced by the phase-folding analysis.

The estimated sensitivity using the Crab Pulsar light-curve and a 1 ms averaged background off-source data sample on the minimum detectable magnitude (U filter) is $m_{U} \sim 13.4$.

\section{Sensitivity to isolated optical pulses: Slewing test}



To test the central pixel's sensitivity to optical flashes, dedicated observations were performed. Central pixel's data were collected during MAGIC's slewing. By using this method, optical flashes (produced by the stars passing by the central pixel field of view - FoV) of known brightness and length were guided into the central pixel. This method allows to experimentally determine the correlation between the maximum voltage of a pulse with the known magnitude of the stars producing them.


During standard operation, MAGIC maximum slewing speed is $\sim$ 4.7 deg/s, so stars passing by the central pixel FoV ($\sim$ 0.1 deg) would produce optical pulses of $\sim$ 20 ms, smeared by the optical point-spread-function (PSF) of the telescope optics. The coordinates of the FoV at a given time are taken directly from MAGIC drive system reports \cite{MAGIC_drive}. The slewing was performed in the azimuthal direction, fixing the zenith angle to the one corresponding to Polaris, in order to test the central pixel saturation and different sub-systems clock matching.

\begin{figure}[h]
\centering
\includegraphics[width=0.8\textwidth]{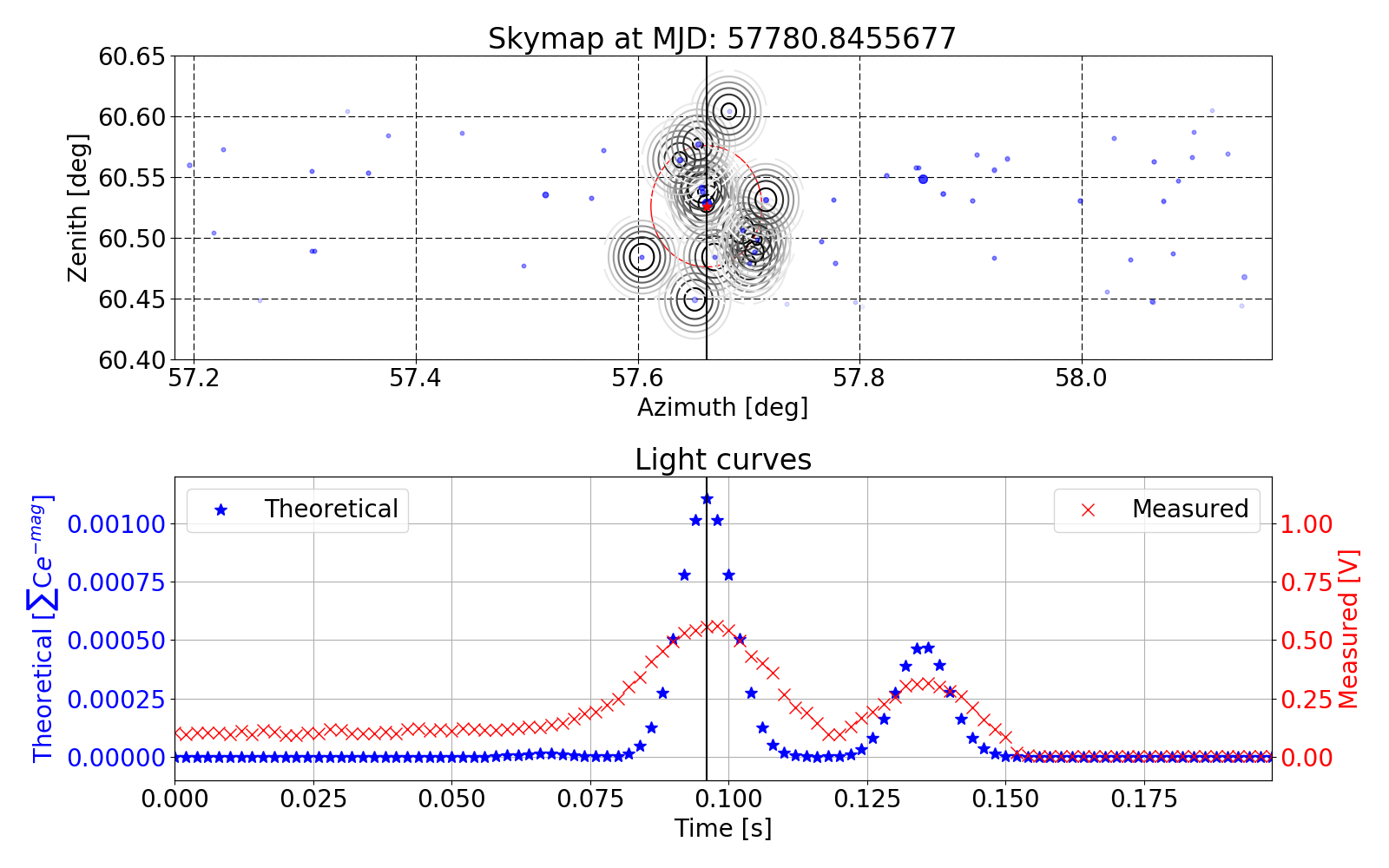}
\caption{Example snapshot of the slewing test. \textit{Top}: Skymap in horizontal coordinates, together with the central pixel field of view (red) taken from the MAGIC drive system reports. Stars position (blue circles) and magnitude (proportional to the marker size) are taken from Vizier catalogs. A Gaussian smearing is applied to the stars within the central pixel's FoV. \textit{Bottom}: Theoretical (blue *) and measured (red $\times$) light curves clearly showing isolated optical pulses when the central pixel passed through several bright stars. The peak heights obtained from fitting a double-Gaussian function to each light curve will be used to obtain the voltage-magnitude relation.}
\label{fig:Slewing test}
\end{figure}

As an example, Fig. \ref{fig:Slewing test} shows a 0.16 s time window during the slewing test. Our aim is to correlate the optical pulses measured by the central pixel with the theoretical ones, calculated by the Gaussian smearing of the the stars magnitude (using \cite{astropy,astroquery, vizier}) within the central pixel's FoV. The width of the Gaussian used is the MAGIC optical PSF. This method will allow a direct measurement of the voltage \emph{vs.} magnitude dependence. The central pixel's sensitivity to isolated optical pulses is determined by extrapolating the fitted curve to the voltage corresponding to a 5 $\sigma$ excess over the background noise level (as defined in section \ref{sec:crab}). 

\begin{figure}[ht]
\centering
\includegraphics[width=0.8\textwidth]{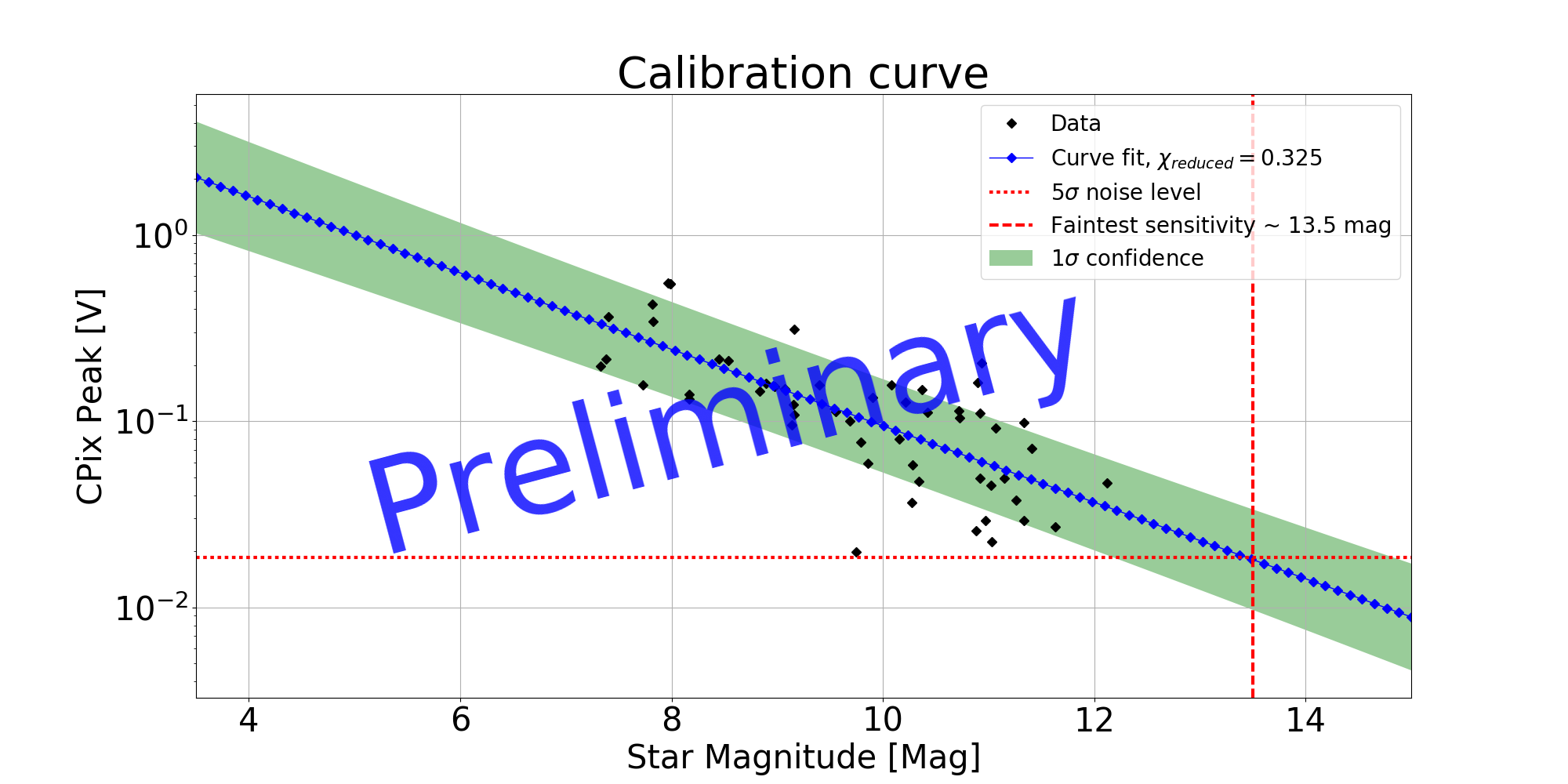}
\caption{Central pixel maximum voltage as a function of the correlated star magnitude of the pulses surviving quality cuts. 
A linear fit together with the 1 $\sigma$ contour are shown. Sensitivity is tested by extrapolating the linear fit to the voltage corresponding to a 5 $\sigma$ excess over the background noise level.}
\label{fig:Calibration curve}
\end{figure}

To eliminate the possibility of uncorrelated pulses, only those with profiles matching the telescope slewing speed were considered. As an additional quality cut, only the amplitudes of consecutively correlated peaks were used as input. The resulting calibration curve is shown in Figure~\ref{fig:Calibration curve}. 

It must be noted that the quality of this measurement is limited by the instabilities of the MAGIC II telescope while slewing. Both the exact pointing position and specially the optical PSF may be unstable during these observations, which might account for the dispersion observed in Fig. \ref{fig:Calibration curve}. Nevertheless, the linear relation between the voltage and expected flux of correlated pulses, and the good agreement with the sensitivity extracted from the Crab Pulsar light-curve seem to demonstrate that the method is working as desired.

\section{Conclusions}

The operating principles of the upgraded MAGIC central pixel, a fine$-$time$-$resolution photometer with multi-channel readout, have been described. MAGIC telescopes are able to simultaneously operate both as VHE and optical telescopes, with excellent sensitivity in the two regimes. After the upgrade, MAGIC is able to detect the Crab optical pulsation in less than 10 s.

The discovery of new fast transient phenomena, such as FRBs, motivates the study of MAGIC sensitivity to short-time-scale (millisecond) isolated flashes. By studying the dispersion measured within off-source data and making use of the well known flux and phaseogram of the Crab Pulsar, we estimated that MAGIC central pixel is able to detect orphan millisecond optical flashes as faint as $\sim$ 13.4 magnitudes (in the U band).

To test this claim and measure the central pixel response to optical flashes of known brightness, we performed dedicated observations collecting central pixel data while slewing. By fitting the voltage of correlated pulses with respect to the magnitude of the stars producing them, and extrapolating this linear relation, the sensitivity to detect a 1 ms optical flash is $m = 13.5 \pm 0.6$.


\section{Acknowledgements}
We would like to thank the IAC for the excellent working conditions at the ORM in La Palma. We acknowledge the financial support of the German BMBF, DFG and MPG, the Italian INFN and INAF, the Swiss National Fund SNF, the European ERDF, the Spanish MINECO, the Japanese JSPS and MEXT, the Croatian CSF, and the Polish MNiSzW.

\bibliographystyle{JHEP}
\bibliography{references}

\providecommand{\href}[2]{#2}\begingroup\raggedright\begin{thebibliography}{10}

\bibitem{FRB_discov}
D.~R. {Lorimer}, M.~{Bailes}, M.~A. {McLaughlin}, D.~J. {Narkevic} and
  F.~{Crawford}, \emph{{A Bright Millisecond Radio Burst of Extragalactic
  Origin}}, \href{http://dx.doi.org/10.1126/science.1147532}{\emph{Science}
  {\bfseries 318} (Nov., 2007) 777},
  [\href{https://arxiv.org/abs/0709.4301}{{\ttfamily 0709.4301}}].

\bibitem{FRB_repeater}
L.~G. {Spitler}, P.~{Scholz}, J.~W.~T. {Hessels}, S.~{Bogdanov}, A.~{Brazier},
  F.~{Camilo} et~al., \emph{{A repeating fast radio burst}},
  \href{http://dx.doi.org/10.1038/nature17168}{\emph{Nature} {\bfseries 531}
  (Mar., 2016) 202--205}, [\href{https://arxiv.org/abs/1603.00581}{{\ttfamily
  1603.00581}}].

\bibitem{aqueye}
C.~{German{\`a}}, L.~{Zampieri}, C.~{Barbieri}, G.~{Naletto}, A.~{{\v C}ade{\v
  z}}, M.~{Calvani} et~al., \emph{{Aqueye optical observations of the Crab
  Nebula pulsar}},
  \href{http://dx.doi.org/10.1051/0004-6361/201118754}{\emph{Astronomy and
  Astrophysics} {\bfseries 548} (Dec., 2012) A47},
  [\href{https://arxiv.org/abs/1210.1796}{{\ttfamily 1210.1796}}].

\bibitem{aqueye+}
L.~{Zampieri}, G.~{Naletto}, C.~{Barbieri}, E.~{Verroi}, M.~{Barbieri},
  G.~{Ceribella} et~al., \emph{{Aqueye+: a new ultrafast single photon counter
  for optical high time resolution astrophysics}},  in \emph{Photon Counting
  Applications 2015}, vol.~9504 of \emph{Proceedings of Science}, p.~95040C,
  May, 2015, \href{https://arxiv.org/abs/1505.07339}{{\ttfamily 1505.07339}},
  \href{http://dx.doi.org/10.1117/12.2179547}{DOI}.

\bibitem{ultracam}
V.~S. {Dhillon}, T.~R. {Marsh}, M.~J. {Stevenson}, D.~C. {Atkinson},
  P.~{Kerry}, P.~T. {Peacocke} et~al., \emph{{ULTRACAM: an ultrafast,
  triple-beam CCD camera for high-speed astrophysics}},
  \href{http://dx.doi.org/10.1111/j.1365-2966.2007.11881.x}{\emph{MNRAS}
  {\bfseries 378} (July, 2007) 825--840},
  [\href{https://arxiv.org/abs/0704.2557}{{\ttfamily 0704.2557}}].

\bibitem{quanteye}
D.~{Dravins}, C.~{Barbieri}, R.~A.~E. {Fosbury}, G.~{Naletto}, R.~{Nilsson},
  T.~{Occhipinti} et~al., \emph{{QuantEYE: The Quantum Optics Instrument for
  OWL}}, {\emph{ArXiv Astrophysics e-prints} (Nov., 2005) },
  [\href{https://arxiv.org/abs/astro-ph/0511027}{{\ttfamily
  astro-ph/0511027}}].

\bibitem{MAGIC_cpix}
F.~Lucarelli, J.~Barrio, P.~Antoranz, M.~Asensio, M.~Camara, J.~Contreras
  et~al., \emph{The central pixel of the magic telescope for optical
  observations},
  \href{http://dx.doi.org/http://dx.doi.org/10.1016/j.nima.2008.03.007}{\emph{Nuclear
  Instruments and Methods in Physics Research} {\bfseries 589} (2008) 415 --
  424}.

\bibitem{veritas_opt}
S.~C. {Griffin}, \emph{{Searching for Fast Optical Transients using a Veritas
  Cherenkov Telescope}},  in \emph{New Horizons in Time Domain Astronomy}
  (E.~{Griffin}, R.~{Hanisch} and R.~{Seaman}, eds.), vol.~285 of \emph{IAU
  Symposium}, pp.~321--323, Apr., 2012,
  \href{https://arxiv.org/abs/1206.6535}{{\ttfamily 1206.6535}},
  \href{http://dx.doi.org/10.1017/S1743921312000932}{DOI}.

\bibitem{hess}
C.~{Deil}, W.~{Domainko}, G.~{Hermann}, A.~C. {Clapson}, A.~{F{\"o}rster},
  C.~{van Eldik} et~al., \emph{{Capability of Cherenkov telescopes to observe
  ultra-fast optical flares}},
  \href{http://dx.doi.org/10.1016/j.astropartphys.2008.12.008}{\emph{Astroparticle
  Physics} {\bfseries 31} (Mar., 2009) 156--162},
  [\href{https://arxiv.org/abs/0812.3966}{{\ttfamily 0812.3966}}].

\bibitem{MAGIC_II}
D.~{Borla Tridon}, F.~{Goebel}, D.~{Fink}, W.~{Haberer}, J.~{Hose}, C.~C. {Hsu}
  et~al., \emph{{Performance of the Camera of the MAGIC II Telescope}},
  {\emph{ArXiv e-prints} (June, 2009) },
  [\href{https://arxiv.org/abs/0906.5448}{{\ttfamily 0906.5448}}].

\bibitem{MII_camera}
f.~{Borla-Tridon}, D. et~al, \emph{{Performance of the Camera of the MAGIC II
  Telescope}}, {\emph{Proceedings of the 31st International Cosmic Rays
  Conference, Lodz, Poland, 2009} (June, 2009) },
  [\href{https://arxiv.org/abs/arXiv:0906.5448}{{\ttfamily arXiv:0906.5448}}].

\bibitem{MAGIC_CP}
M.~{Lopez}, N.~{Otte}, M.~{Rissi}, T.~{Schweizer}, M.~{Shayduk}, S.~{Klepser}
  et~al., \emph{{Detection of the crab pulsar with MAGIC}}, {\emph{ArXiv
  e-prints} (July, 2009) }, [\href{https://arxiv.org/abs/0907.0832}{{\ttfamily
  0907.0832}}].

\bibitem{Bessel}
M.~S. {Bessell}, \emph{{UBVRI photometry. II - The Cousins VRI system, its
  temperature and absolute flux calibration, and relevance for two-dimensional
  photometry}}, \href{http://dx.doi.org/10.1086/130542}{\emph{Publications of
  the ASP} {\bfseries 91} (Oct., 1979) 589--607}.

\bibitem{Sollerman}
J.~{Sollerman}, P.~{Lundqvist}, D.~{Lindler}, R.~A. {Chevalier}, C.~{Fransson},
  T.~R. {Gull} et~al., \emph{{Observations of the Crab Nebula and Its Pulsar in
  the Far-Ultraviolet and in the Optical}},
  \href{http://dx.doi.org/10.1086/309062}{\emph{Astrophysical Journal}
  {\bfseries 537} (July, 2000) 861--874},
  [\href{https://arxiv.org/abs/astro-ph/0002374}{{\ttfamily
  astro-ph/0002374}}].

\bibitem{MAGIC_drive}
T.~{Bretz}, D.~{Dorner}, R.~M. {Wagner} and P.~{Sawallisch}, \emph{{The drive
  system of the major atmospheric gamma-ray imaging Cherenkov telescope}},
  \href{http://dx.doi.org/10.1016/j.astropartphys.2008.12.001}{\emph{Astroparticle
  Physics} {\bfseries 31} (Mar., 2009) 92--101},
  [\href{https://arxiv.org/abs/0810.4593}{{\ttfamily 0810.4593}}].

\bibitem{astropy}
{Astropy Collaboration}, T.~P. {Robitaille}, E.~J. {Tollerud}, P.~{Greenfield},
  M.~{Droettboom}, E.~{Bray} et~al., \emph{{Astropy: A community Python package
  for astronomy}},
  \href{http://dx.doi.org/10.1051/0004-6361/201322068}{\emph{Astronomy and
  Astrophysics} {\bfseries 558} (Oct., 2013) A33},
  [\href{https://arxiv.org/abs/1307.6212}{{\ttfamily 1307.6212}}].

\bibitem{astroquery}
A.~Ginsburg, T.~Robitaille, M.~Parikh, C.~Deil, J.~Mirocha, J.~Woillez et~al.,
  \emph{{Astroquery v0.1}}, .

\bibitem{vizier}
F.~{Ochsenbein}, P.~{Bauer} and J.~{Marcout}, \emph{{The VizieR database of
  astronomical catalogues}},
  \href{http://dx.doi.org/10.1051/aas:2000169}{\emph{Astronomy and
  Astrophysics, Supplement} {\bfseries 143} (Apr., 2000) 23--32},
  [\href{https://arxiv.org/abs/astro-ph/0002122}{{\ttfamily
  astro-ph/0002122}}].

\end{thebibliography}\endgroup

\end{document}